%% file: main.tex
\documentclass[a4paper, 10pt, twocolumn]{article} 
\input{gamma26-template}

\title{
    From photons to neutrinos : Advancing multi-messenger modeling with Gammapy
}

\author{
	\authorstyle{J.~Albinet\textcolor{GammaDarkGreen}{\textsuperscript{1}}, L.~Heckmann\textcolor{GammaDarkGreen}{\textsuperscript{1}}, M.~Cerruti\textcolor{GammaDarkGreen}{\textsuperscript{1}}, B.~Khélifi\textcolor{GammaDarkGreen}{\textsuperscript{1}} \onbehalf{}}
	\newline\newline
    {\footnotesize
        \textsuperscript{\color{GammaDarkGreen}1}\institution{Université Paris Cité, CNRS, Astroparticule et Cosmologie, F-75013 Paris, France.}
    }\\
}

\date{\today}
\newcommand{\website}{\href{https://plan.events.mpg.de/event/543/}{Proceedings of the 9th Heidelberg Gamma-Ray Symposium}} 

\begin{document}
\sansmath\maketitle
\thispagestyle{fancy}

\lettrineabstract{In recent years, several astrophysical objects have been identified as potential multi-messenger photon-neutrino emitters. New instruments dedicated to the study of these two messengers have been and are still being built, achieving unprecedented sensitivity. This observational improvement increases the need for more sophisticated and precise analysis tools. Multi-wavelength and multi-messenger data fitting is crucial to constrain astrophysical models of particle acceleration and radiation mechanisms at work in these extreme environments.
In this contribution, we present a novel framework dedicated to multi-wavelength and multi-messenger modeling, built within the open-source software Gammapy. The framework enables data analysis from optical up to very-high-energy gamma rays, and fitting physical models directly to instrumental counts in these wavebands, without the need of computing flux points.
Focusing on blazars as a first example of a multi-messenger candidate, we showcase the performance of this tool in constraining multi-wavelength emission models, and how it reduces fitting biases. We show the results of a leptonic model that can be fitted on the fly, due to the short computing time needed. For more computing-intensive emission codes, we explore a grid-fitting approach. We evaluate the performance of the new fitting tool  and present its first application to leptonic modeling of blazars. By fitting directly on photon and neutrino events, this novel approach will enable, for the first time, statistically sound constraints on the physics of multi-messenger sources.}

\section{Introduction}

The study of extreme particle accelerators requires collecting and analyzing data from a large number of instruments, covering a large part of the photon spectrum, or combining photons with additional messengers such as neutrinos or gravitational waves.  
Current approaches for multi-wavelength and multi-messenger analyses typically involve the computation of flux points estimated in each waveband, followed by fitting a physical model with $\chi^2$ statistics. This method has, however, several limitations. First, flux points are derived with the assumption that the flux in each energy band follows a functional form (such as a Power Law or a Log Parabola), which is different from the model that will later be fitted to these flux points, introducing a bias. Additionally, upper limits (ULs) are often only used as strict thresholds, to ensure that the model does not overshoot them.\\

The software Gammapy \cite{gammapy:2023} is the core library for the high-level analysis of CTAO data. Its extension, gammapy-mwl \footnote{\url{https://github.com/gammapy/gammapy-mwl}}, enables the analysis of data ranging from optical to gamma-rays. We built a framework based on this extension with the version 2.0 of Gammapy \cite{acero_2025_17814297} that allows for statistically sound multi-wavelength fitting of arbitrary models through forward-folding to instrumental counts. An approach using functional forms fitting on multi-wavelength data has been presented in \cite{Nievas_Rosillo_2025}. Because this framework operates directly on events/counts, it inherently accounts for ULs and does not introduce any bias from flux-point estimation. The performance of this tool is presented in the following sections.

\begin{figure*}[t!]
    \centering
    \includegraphics[width=0.4\linewidth]{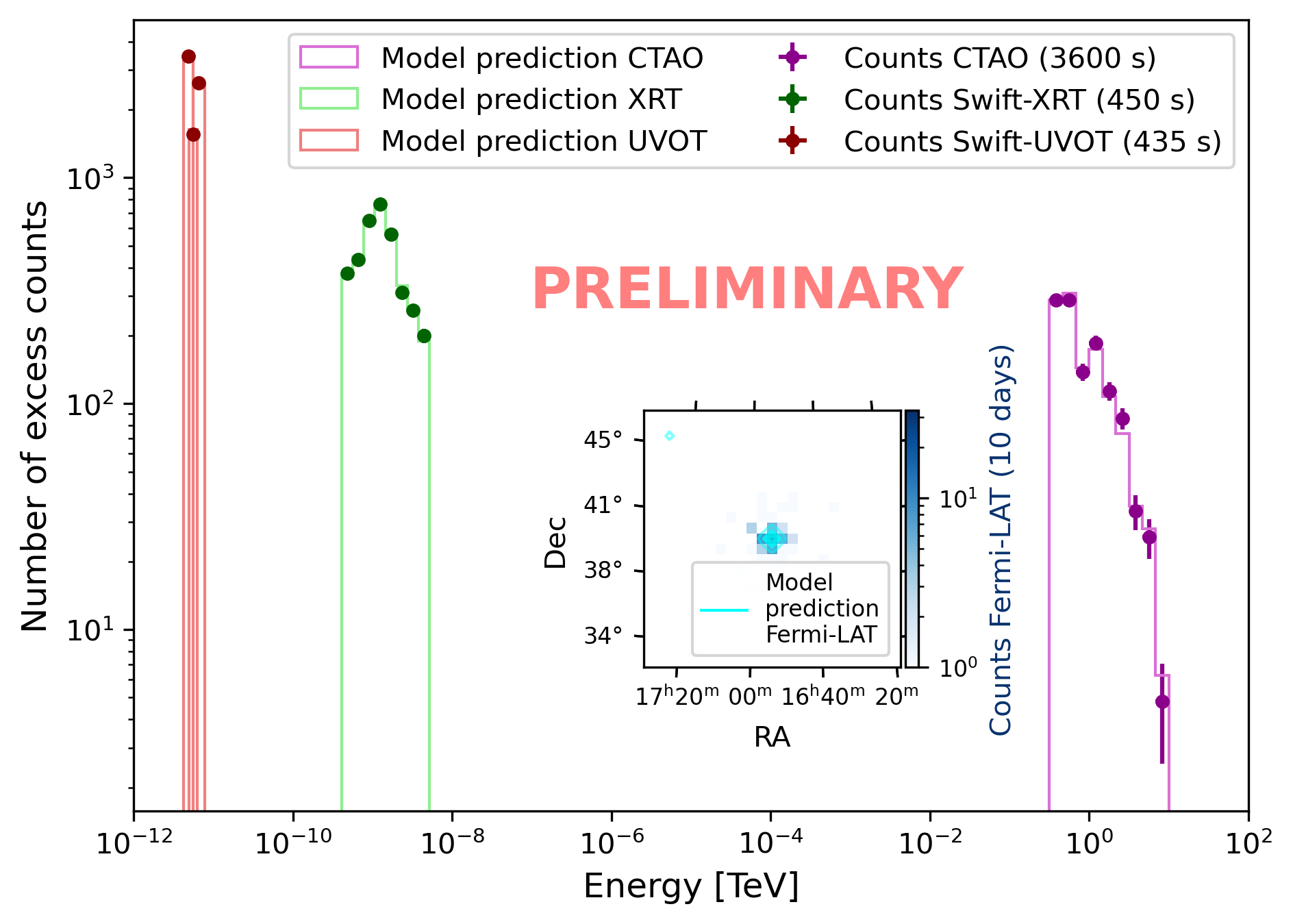}
    \includegraphics[width=0.44\linewidth]{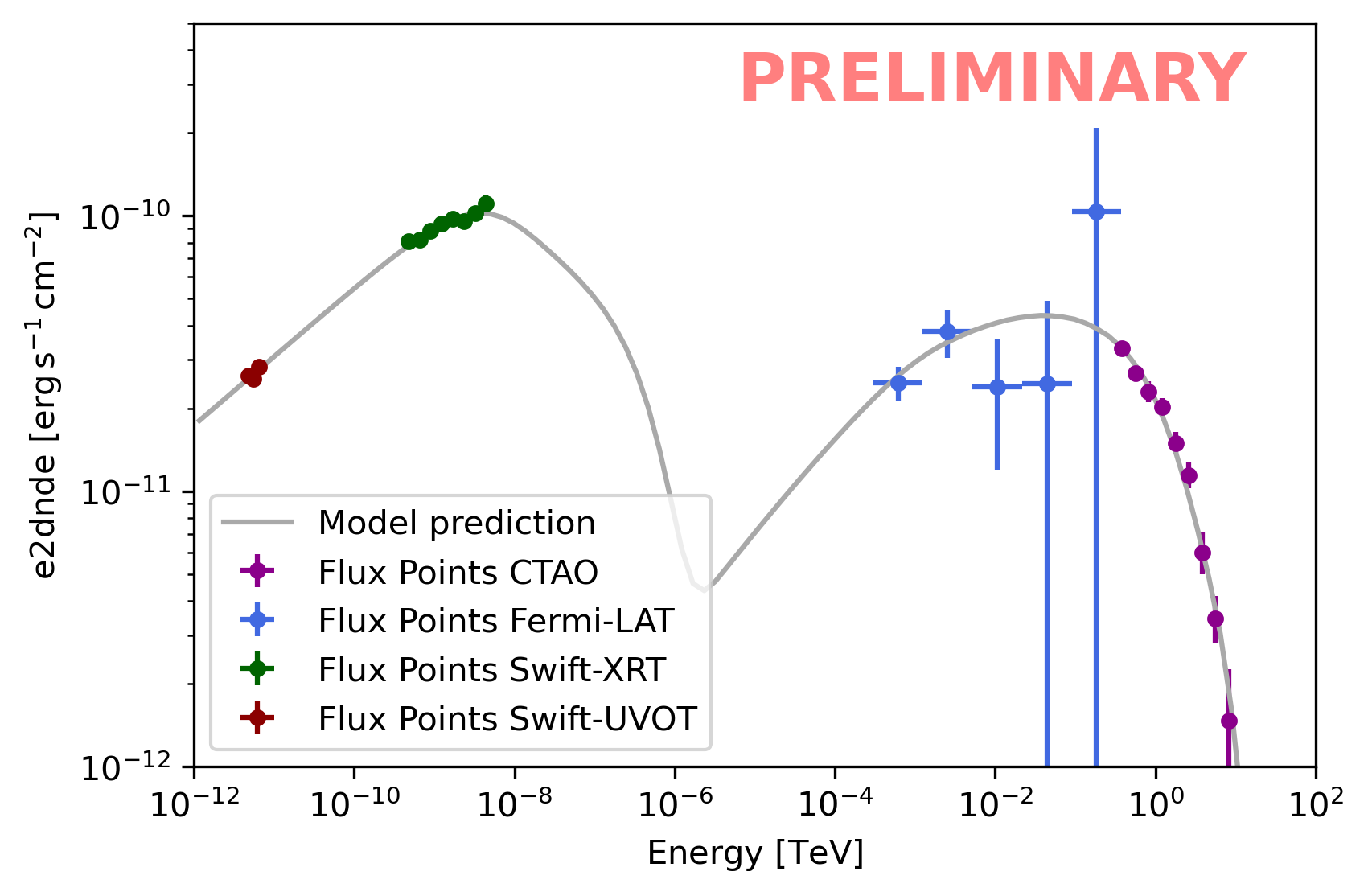}
    \caption{\textit{Left:} simulated binned counts for SWIFT-UVOT/XRT and CTAO and simulated Fermi-LAT count map. \textit{Right:} flux points for each instrument and simulated SSC model (solid line).}
    \label{fig:events}
\end{figure*}

\section{An application: simulating blazar emission}

We created two types of datasets to probe the performance of a fit to counts compared to a standard fit to flux points. We simulated the count datasets with real instrument response functions (IRFs) of UVOT, XRT and LAT and the following simulated IRF of CTAO \path{Prod5-North-40deg-AverageAz-4LSTs09MSTs.18000s-v0.1.fits}\cite{cherenkov_telescope_array_observatory_2021_5499840}. The assumed spectral model was a Synchrotron Self-Compton (SSC) model of a blazar, simulated using the Gammapy plugin of the JetSeT code \cite{2020AT,2011AT,2009AT}. The model includes optical reddening, absorption by neutral hydrogen in soft-X-rays and absorption by pair-production on the extragalactic background light. As a first test, we simulated the spectral energy distribution of a typical high- to extreme-high-frequency-peaked BL Lac object, using model parameters close to the ones used in the modeling of the well-known blazar Mrk 501 \cite{Mrk501}.
The simulated count datasets were created from UVOT and XRT data from a single Swift pointing of 450 s, Fermi/LAT data integrated over 10 days, and a 1-hour simulation of CTAO observations (using the North array). 
We generated flux points for each instrument separately, and merged them into a flux-point Gammapy dataset, as it is the current standard in multi-wavelength studies. Within each individual band, we adopted as spectral model a power-law function, or a log parabola if it was favored by more than 3 $\sigma$ over the power law.  An example of the simulated count dataset, and the flux-point dataset is shown in the left and right plots of Fig. \ref{fig:events}, respectively.

\section{Performance of the two algorithms}

We simulated 100 different pairs of count and flux-point datasets to study the average properties of the new framework. We then fit each dataset with the same starting parameters, using the SSC JetSeT plugin in Gammapy.
As we expect convergence to be more difficult to achieve with a higher number of free parameters, we freed the SSC parameters in the following order: \( B \), \( N \), \( \gamma_{\text{break}} \), \( p_1 \), and \( p \). The simulated values are \( B = 0.1\,\text{G} \), \( N = 5\,\,\text{cm}^{-3} \), \( \gamma_{\text{break}} = 5 \times 10^5 \), \( p_1 = 3.5 \), and \( p = 2.5 \).
Furthermore, the count datasets possess additional degrees of freedom compared to the flux-point datasets because of free normalization of other sources in the Fermi field of view. 

The minimization within Gammapy was done using iminuit \cite{iminuit}. 
Two approaches were tested for initializing the parameters:
\begin{itemize}
    \item \textbf{Good initial guess:}
    Free parameters among those mentioned before are initialized with \( B = 0.05\,\text{G} \), \( N = 10\,\,\text{cm}^{-3} \), \( \gamma_{\text{break}} = 2 \times 10^5 \), \( p_1 = 3.2 \), and \( p = 2.2 \), while frozen parameters are set to their simulated values. This approach mimics a pre-fit of a model by eye, which we would typically attempt in order to estimate reasonable starting parameters before running a fit. In this case, all 100 fits for both flux-point and count datasets share the same starting parameters.
    \item \textbf{Random initial guess :}
    Starting parameters are drawn randomly following a uniform distribution. Here, each pair of flux-point and count datasets shares the same random starting parameters, but these differ across pairs.
\end{itemize}

\begin{figure*}
    \centering
    \includegraphics[width=0.4\linewidth]{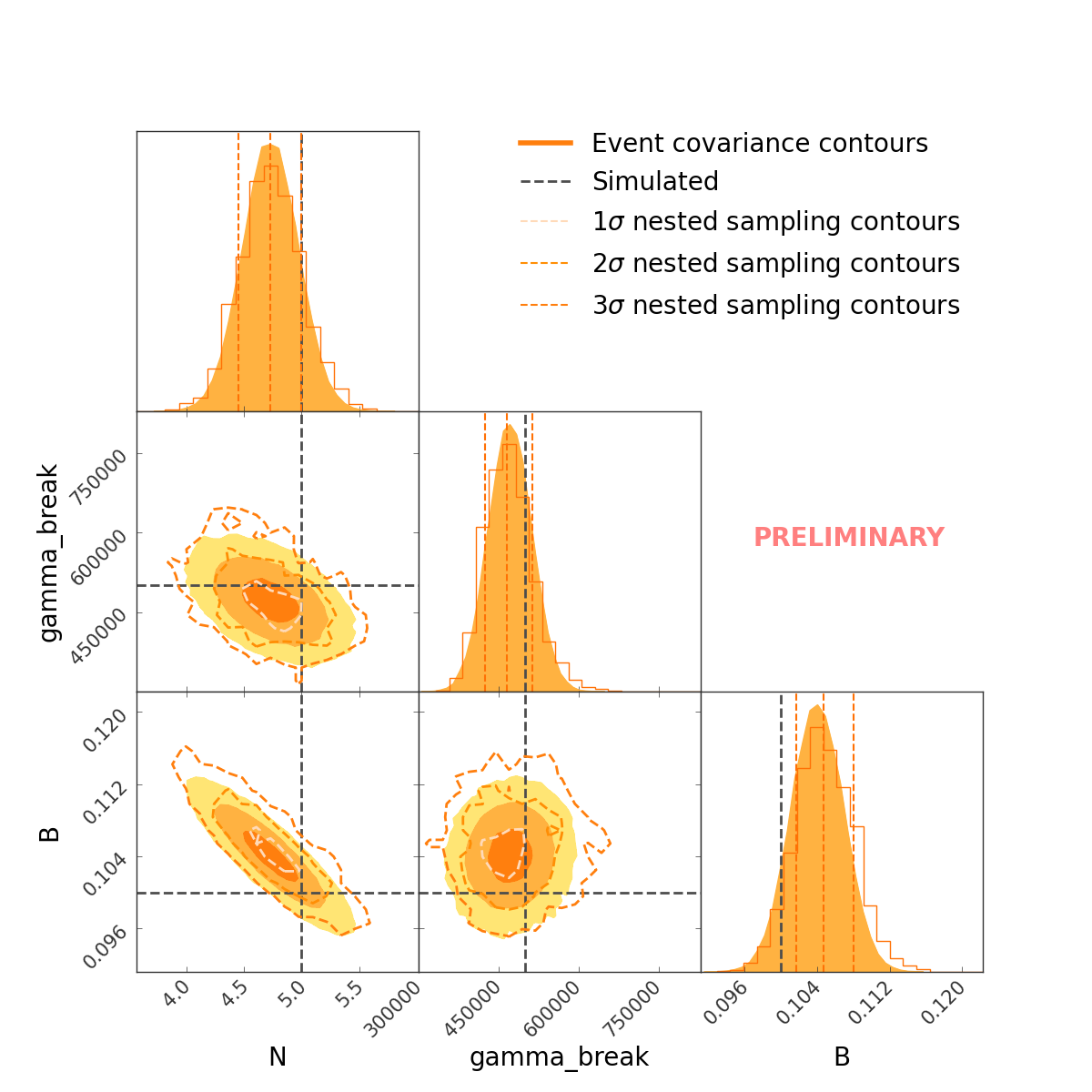}
    \includegraphics[width=0.4\linewidth]{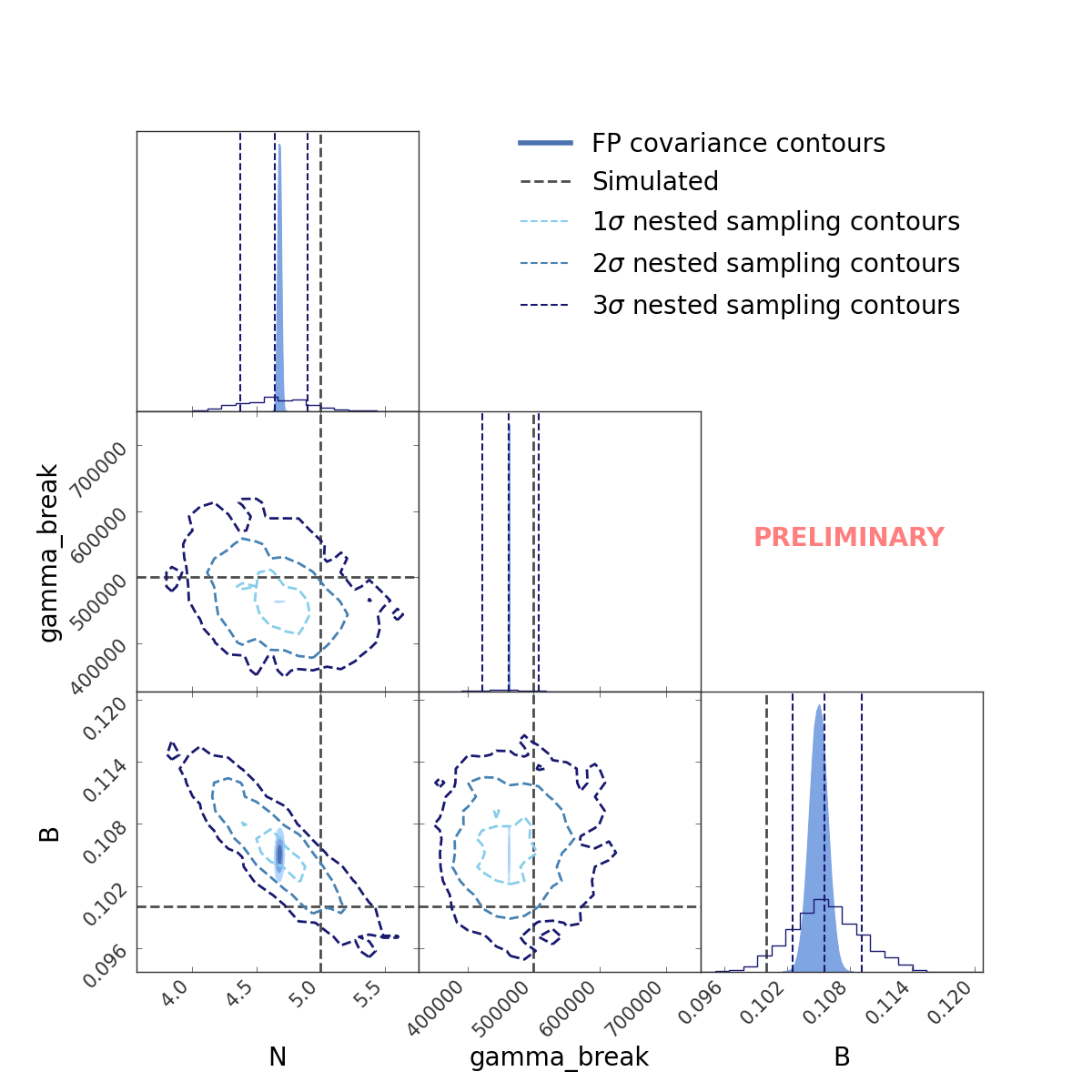}
    
    \caption{\textit{Left:} $1-3~\sigma$ iminuit covariance contours for the count dataset (solid orange contours) and nested sampling equivalents (dashed orange lines). The 1D histograms and nested sampling posteriors for each parameter are represented on the diagonal plots. The simulated values of the parameters are shown in gray dashed lines. \textit{Right:} $1-3~\sigma$ iminuit covariance contours for the flux-point dataset (plain blue contours) and nested sampling equivalents (dashed blue lines). The 1D histograms and nested sampling posteriors for each parameter are represented on the diagonal plots. The simulated values of the parameters are shown in gray dashed lines.}
    \label{fig:events_fit}
\end{figure*}

Additionally, we derived posterior probability distributions assuming uniform priors and the Bayesian evidence with the nested sampling Monte Carlo algorithm MLFriends \cite{2016Buchner,2019Buchner} using the UltraNest\footnote{\url{https://johannesbuchner.github.io/UltraNest/}} package \cite{2021Buchner}. We then compared the contours extracted from the covariance matrix, estimated with iminuit strategy 1, with the nested sampling posterior distributions. An example of a convergent fit of a single realization is shown in the left and right plots of Fig. \ref{fig:events_fit},  for the count dataset and the flux-point dataset, respectively.\\

The fitted physical parameters yield very similar values whether we use the flux-point or count dataset. While the nested sampling posteriors align well with the covariance contours for the count dataset, the covariance contours for the flux-point dataset are significantly smaller than the nested sampling posteriors. This discrepancy suggests that errors are greatly underestimated when fitting to flux points.
We can further support the former statement by looking at the number of realizations for which the fitted parameters are within a $5\sigma$ interval around the simulated parameters. We compute this interval through the Mahalanobis distance (\ref{Mahalanobis}).
\begin{equation}
    D(u,v,\mathrm{V}) = \sqrt{(u-v)\times \mathrm{V}^{-1}\times (u-v)^{\mathrm{T}}}
    \label{Mahalanobis}
\end{equation}

In this equation, u and v represent the vectors of fitted and simulated parameters, and V is the covariance matrix. \\

\begin{figure}
    \centering
    \includegraphics[width=0.8\linewidth]{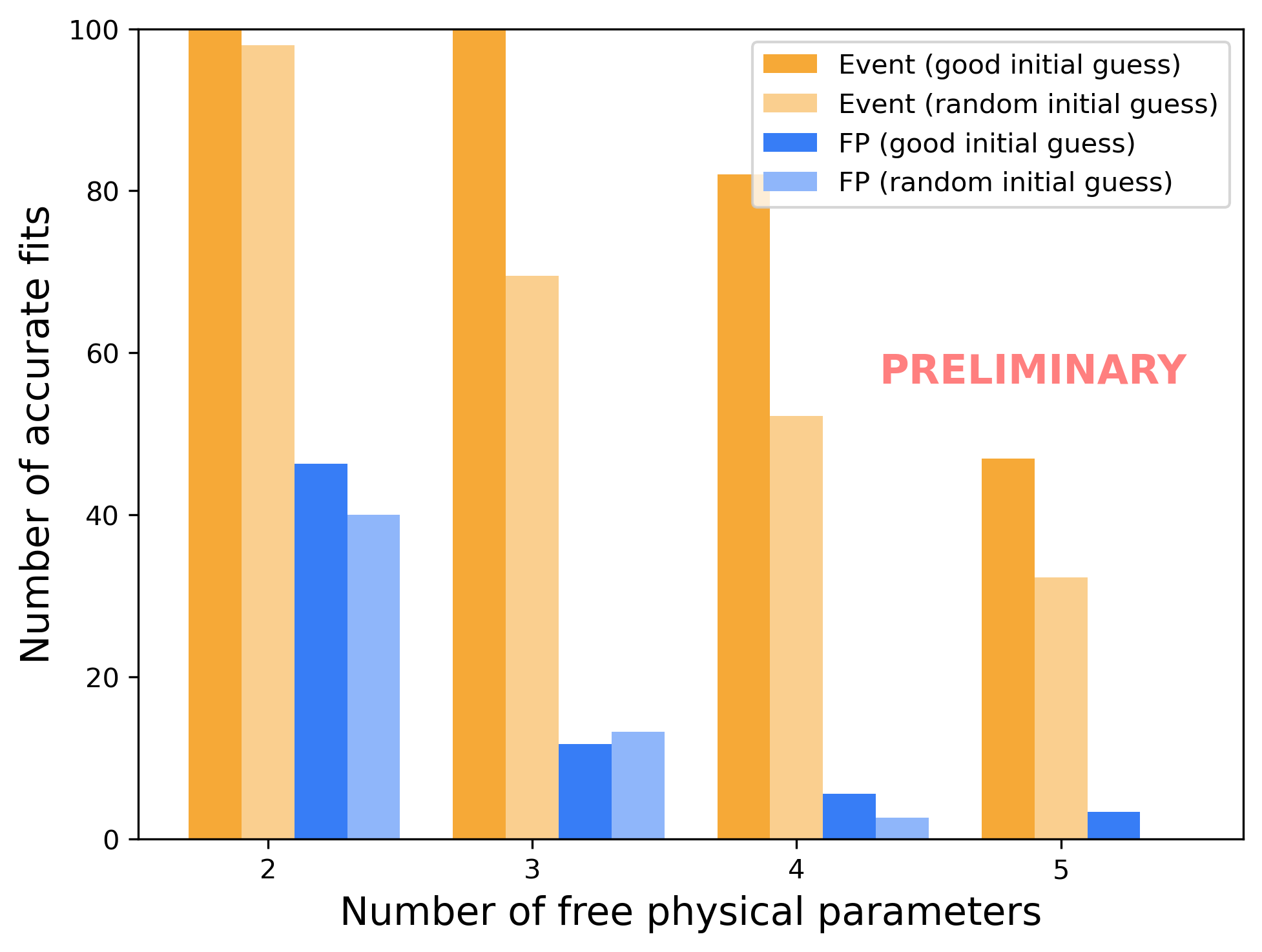}
    \caption{Number of convergent fits with reliable uncertainties and within the 5$\sigma$ interval around the simulated values for the flux-point and count datasets.}
    \label{fig:histogram}
\end{figure}

Fitting to counts significantly increases convergence compared to fitting to flux points. Any conclusion on the quality of the fits is however non-trivial. To address the latter point, we evaluate how many of the 100 realizations produced convergent fits with reliable uncertainties and a Mahalanobis distance of less than 5 between the fitted and simulated parameters, which we call "accurate fits".
The results of these two approaches are shown in Fig. \ref{fig:histogram}. The orange and blue bars respectively represent the count and flux-point datasets. Darker bars correspond to fits initialized with a good guess, while lighter bars represent fits initialized with random parameters. The reliability of the fit to flux points is low even with a low number of free physical parameters. On the other hand, the fit to counts is more robust even with a high number of free parameters. In addition, starting with a good guess of the initial parameters matters, especially when several physical parameters are free.\\

In terms of computing time, the fit to counts is not significantly longer than the fit to flux points. Indeed, the standard fit to flux points with 3 free physical parameters takes $\sim 5~\mathrm{min}$, while the standard fit on counts takes $\sim 10~\mathrm{min}$. The nested sampling, which improves the uncertainty on the parameters for the fit to flux points, is much more computationally intensive, and requires over $3~\mathrm{h}$ for both the flux point and count datasets.\\

\section{Conclusion and outlook}

We developed a novel framework within the open-source Gammapy software, dedicated to unbiased multi-wavelength and multi-messenger fitting. This framework enables fast and robust fitting of any physical model directly on instrumental counts. First applications show that it not only improves convergence but also provides statistically accurate errors on the physical parameters. Nested sampling, whether used on counts or flux points, remains the most accurate method to estimate parameter values and uncertainties, at the expense of a much longer computing time. However, multi-wavelength fitting with iminuit to counts provides a much faster convergence and solid estimate of uncertainties. The uncertainties on the fitted parameters estimated from the iminuit fit to flux points cannot be trusted. 
While our application focuses on a leptonic model, this tool can be used on (lepto-)hadronic models as well. Neutrino analysis is already compatible with Gammapy \footnote{\url{https://github.com/antoniocondorelli/gammapy-acme}}$^{,}$\footnote{\url{https://agenda.infn.it/event/47822/contributions/293499/attachments/153411/235311/gammapy_hands_on-2-3-2.pdf}} and will soon be officially implemented allowing for statistically sound multi-messenger fits. Since hadronic fits are computationally intensive, we are exploring the use of pre-computed grids and machine learning solutions to reduce the computing time and make the fitting algorithm attractive for a large variety of astrophysical applications.

\section{Acknowledgments}
This work benefited from the support of the project COCOA-NuGETs ANR-23-CE31-0026 of the French National Research Agency (ANR). This work made use of Gammapy \cite{gammapy:2023}, a community-developed Python package. The Gammapy team acknowledges all Gammapy past and current contributors, as well as all contributors of the main Gammapy dependency libraries: \href{https://numpy.org/}{NumPy}, \href{https://scipy.org/}{SciPy}, \href{http://www.astropy.org}{Astropy}, \href{https://astropy-regions.readthedocs.io/}{Astropy Regions}, \href{https://scikit-hep.org/iminuit/}{iminuit}, \href{https://matplotlib.org/}{Matplotlib}.

\printbibliography[title={References}]


\end{document}

%% file: gamma26-template.tex
\usepackage[british]{babel}
\usepackage{microtype}
\usepackage{amsmath,amsfonts,amsthm}
\usepackage[svgnames,table]{xcolor}
\definecolor{GammaDarkGreen}{rgb}{0.113,0.278,0.2745}
\definecolor{GammaCharcoal}{rgb}{0.0706, 0.0706, 0.0706}
\definecolor{GammaRed}{HTML}{B71325}
\definecolor{GammaGreen}{HTML}{004846}
\usepackage{booktabs}
\usepackage{lastpage}
\usepackage{graphicx}
\usepackage{enumitem}
\setlist{leftmargin=*,itemsep=0.4ex,parsep=0ex,topsep=0.8ex,partopsep=0ex}
\setlist[itemize]{label=\color{GammaRed}$\triangleright$}
\usepackage{sectsty}
\chapterfont{\normalfont\Large\color{GammaRed}}
\sectionfont{\normalfont\Large\color{GammaRed}}
\subsectionfont{\normalfont\large\color{GammaCharcoal}}
\subsubsectionfont{\normalfont\normalsize\color{GammaGreen}}

\usepackage{subfig}

\usepackage{caption}
\DeclareCaptionLabelFormat{abbrevfig}{\color{GammaRed}Fig.\ #2}
\DeclareCaptionLabelFormat{abbrevtab}{\color{GammaRed}Tab.\ #2}
\usepackage[pdftex]{hyperref}
\pdftrailerid{}
\hypersetup{colorlinks=true,linkcolor=GammaDarkGreen,citecolor=GammaDarkGreen,filecolor=GammaDarkGreen,urlcolor=GammaDarkGreen}

\addto\extrasenglish{%
}

\usepackage{geometry}
\usepackage[utf8]{inputenc}
\usepackage[T1]{fontenc}
\IfFileExists{nimbusmononarrow.sty}{\usepackage{nimbusmononarrow}}{}
\usepackage{tgheros}
\usepackage{sansmath}
\usepackage{fancyhdr}
\usepackage{titling}

\newcommand{\authorstyle}[1]{{\large\color{GammaCharcoal}#1}}
\newcommand{\institution}[1]{{\footnotesize\itshape\color{GammaDarkGreen}#1\/}}
\newcommand{\onbehalf}[1]{{\large\color{GammaCharcoal}#1}}

\pretitle{%
    \noindent
    \begin{minipage}[t]{0.63\textwidth}\vspace{20pt}
        \LARGE\color{GammaRed}
}

\posttitle{%
    \end{minipage}%
    \hfill
    \begin{minipage}[t]{0.33\textwidth}\vspace{-50pt}
        \raggedleft
        \includegraphics[width=\linewidth]{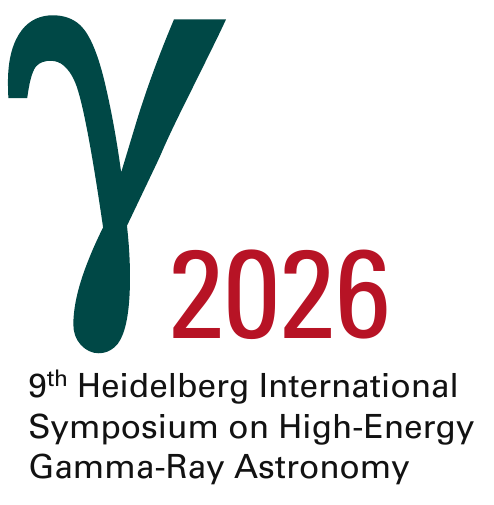}
    \end{minipage}%
}

\preauthor{\color{GammaDarkGreen}}
\postauthor{}
\predate{\begin{flushleft}\large\color{GammaRed}}
\postdate{\par\end{flushleft}}

\usepackage{lettrine}
\usepackage{xstring}
\newcommand{\initial}[1]{\lettrine[lines=2,findent=4pt,nindent=0pt]{\color{GammaRed}\emph{#1}}{}}
\newcommand{\lettrineabstract}[1]{\StrLeft{#1}{1}[\firstletter]\initial{\firstletter}\emph{{\StrGobbleLeft{#1}{1}}}}

\usepackage[backend=bibtex,style=numeric-comp,sorting=none,giveninits=true,citestyle=numeric]{biblatex}
\usepackage[autostyle=true]{csquotes}